\documentclass[10pt,english]{article}
\usepackage{times}
\usepackage[T1]{fontenc}
\usepackage[latin1]{inputenc}
\usepackage{amsmath}
\usepackage{amssymb}

\makeatletter

\newcommand{\noun}[1]{\textsc{#1}}

\usepackage{slashed}

\usepackage{babel}
\makeatother
\begin{document}

\title{\textbf{CRITICAL POINT IN A 3-3-1 MODEL WITH RIGHT-HANDED NEUTRINOS}}

\author{\noun{ADRIAN} PALCU}

\date{\emph{Department of Theoretical and Computational Physics - West
University of Timi\c{s}oara, V. P\^{a}rvan Ave. 4, RO - 300223 Romania}}

\maketitle
\begin{abstract}
The boson mass spectrum of a 3-3-1 model with right-handed neutrinos
is investigated by tuning a unique free parameter whitin the exact
algebraical approach for solving gauge models with high symmetries.
A very strange coincidence of the masses in both the neutral and the
charged boson sectors can occur at a not very high breaking scale.
This could explain why the masses of the new bosons have not been
exactly weighted in the laboratories by date. In order to have good
phenomenological behaviour in the neutrino sector, one can resort
to either the conservation of the global lepton number $L_{e}-L_{\mu}-L_{\tau}$
or see-saw mechanism. 


\end{abstract}

\section{Introduction}

In this brief report we would like to emphasize the fact that in the
3-3-1 gauge model with right-handed neutrinos, a critical point could
well occur at a not very high breaking scale. All the Particle Data
\cite{key-1} suggest that the neutral boson $Z^{\prime}$ of any
extention of the Standard Model (SM) has to be heavier than neutral
boson of the SM. However, the case when the ''old'' neutral boson
screens the ''new'' one - namely, when their masses are identical
- seems to be ignored. A very suitable manner to investigate this
issue is supplied by the exact algebraical approach for solving gauge
theories with high symmetries proposed several years ago by Cot\u{a}escu
\cite{key-2} and developed by the author in a series of recent papers
\cite{key-3} - \cite{key-7} on the 3-3-1 model with right-handed
neutrinos. For some interesting details of the phenomenology in such
models, the reader is reffered to Refs. \cite{key-8} - \cite{key-35}. 

The results regarding the exact boson mass spectrum - presented hereafter
- are amazing and they look plausible from phenomenological viewpoint.
There is a critical point in the model where all the neutral bosons
of the theory $Z$, $Z^{\prime}$ and $Y$ gain the same mass! Moreover,
at the same time the charged bosons $W^{\pm}$ and $X^{\pm}$ are
indistinguishable on the mass reason too. This special feature seems
to explain why the new bosons were not yet exactly ''weighted''
in laboratory. Only their lower energy bounds are suggested in the
literature (see on this issue Ref. \cite{key-1} and references therein). 

The paper is organized as follows. Section 2 briefly reviews the main
results of the theoretical method employed to solve the particular
3-3-1 model with right-handed neutrinos, so the boson mass spectrum
- depending on the sole free parameter $a$ - is given. Section 3
analyses the circumstances under which the critical point occurs and
estimates the phenomenological implications of the particular value
of the parameter $a$ that ensures this behaviour. Last section is
devoted to our conclusions.

\section{Boson Mass Spectrum}

The particle content of the $SU(3)_{c}\otimes SU(3)_{L}\otimes U(1)_{Y}$
gauge model with right-handed neutrinos, under consideration here
\cite{key-8} - \cite{key-35}, is:

\paragraph{Lepton families}

\begin{equation}
\begin{array}{ccccc}
f_{\alpha L}=\left(\begin{array}{c}
\nu_{\alpha}^{c}\\
\nu_{\alpha}\\
e_{\alpha}\end{array}\right)_{L}\sim(\mathbf{1,3},-1/3) &  &  &  & \left(e_{\alpha L}\right)^{c}\sim(\mathbf{1},\mathbf{1},-1)\end{array}\label{Eq.1}\end{equation}

\paragraph{Quark families}

\begin{equation}
\begin{array}{ccc}
Q_{iL}=\left(\begin{array}{c}
D_{i}\\
-d_{i}\\
u_{i}\end{array}\right)_{L}\sim(\mathbf{3,3^{*}},0) &  & Q_{3L}=\left(\begin{array}{c}
T\\
t\\
b\end{array}\right)_{L}\sim(\mathbf{3},\mathbf{3},-1/3)\end{array}\label{Eq.2}\end{equation}

\begin{equation}
\begin{array}{ccc}
(b_{L})^{c},(d_{iL})^{c}\sim(\mathbf{3},\mathbf{1},-1/3) &  & (t_{L})^{c},(u_{iL})^{c}\sim(\mathbf{3},\mathbf{1},+2/3)\end{array}\label{Eq.3}\end{equation}

\begin{equation}
\begin{array}{ccccccccc}
(T_{L})^{c}\sim(\mathbf{3,1},+2/3) &  &  &  &  &  &  &  & (D_{iL})^{c}\sim(\mathbf{3,1},-1/3)\end{array}\label{Eq.4}\end{equation}
with $i=1,2$. The numbers in paranthesis denote - in a self-explanatory
notation - the representations and the charecters with respect to
each group involved in the theory. 

With these representations this particular 3-3-1 model stands anomaly
free, as one can easily check out by using little algebra. Note that,
although all the anomalies cancel by an interplay between familes,
each family still remains anomalous by itself. 

These representations can be achieved starting with the general method
\cite{key-2} of exactly solving gauge models with high symmetries
by just chosing an appropriate set of parameters (for certain details
of dealing with the algebraical procedure and the special parametrisation
involved here, the reader is referred to Ref. \cite{key-7}). They
are:

\begin{equation}
e,\theta_{W},\nu_{0}=0,\nu_{1}=0,\nu_{2}=1\label{Eq.5}\end{equation}
established by experimental arguments ($e,\theta_{W}$) \cite{key-1}
or by internal reasons of the general method ($\nu_{i}$) \cite{key-2}. 

Along with the above parameters, one must add some new ones - as they
determine the Higgs sector of the model - grouped in a parameter matrix
which reads:

\begin{equation}
\eta^{2}=(1-\eta_{0}^{2})Diag\left[1-a,\frac{1}{2}(a+b),,\frac{1}{2}(a-b)\right]\label{Eq.6}\end{equation}
where, for the moment, $a$ and $b$ are arbitrary non-vanishing real
parameters that ensure the condition $Tr\left(\eta^{2}\right)=(1-\eta_{0}^{2})$.
At the same time, $\eta_{0},a\in[0,1)$. 

They will determine, after the spontaneous symmetry breakdown (SSB)
- which takes place up to the universal residual $U(1)_{em}$ one
- a non-degenerate boson mass. The exact expressions of the boson
masses are given by the Eqs. (53) - (55) in Ref. \cite{key-2}, namely

\begin{equation}
M_{i}^{j}=\frac{1}{2}g\left\langle \phi\right\rangle \sqrt{\left[\left(\eta^{(i)}\right)^{2}+\left(\eta^{(j)}\right)^{2}\right]}\label{Eq.7}\end{equation}
for the non-diagonal gauge bosons which usually are charged but, as
one can easily observe in the 3-3-1 model under consideration here,
one of them must be neutral, and

\begin{equation}
\left(M^{2}\right)_{ij}=\left\langle \phi\right\rangle ^{2}Tr\left(B_{i}B_{j}\right)\label{Eq.8}\end{equation}
with \begin{equation}
B_{i}=g\left[D_{i}+\nu_{i}\left(D\nu\right)\frac{1-\cos\theta}{\cos\theta}\right]\eta\label{Eq.9}\end{equation}
for the diagonal bosons of the model. The angle $\theta$ is the rotation
angle around the versor $\nu$ orthogonal to the electromagnetic direction
in the parameter space \cite{key-2}. The versor condition holds $\nu_{i}\nu^{i}=1.$ 

Since the electro-weak sector of the model is described now by the
chiral gauge group $SU(3)_{L}\otimes U(1)_{Y}$, the two diagonal
generators $D_{1}$ and $D_{2}$ ($D$s - stands for the Hermitian
diagonal generators of the Cartan subalgebra) in the fundamental representation
of $SU(3)_{L}$ are: $D_{1}=T_{3}$ and $D_{2}=T_{8}$ - connected
to the Gell-Mann matrices in the manner $T_{a}=\lambda_{a}/2$ - and
$D_{0}=I$ for the chiral hypercharge. 

In the 3-3-1 model under consideration here, the relation between
$\theta$ in the general method \cite{key-2} and the Weinberg angle
$\theta_{W}$ from SM was established \cite{key-3,key-7} and it is 

\begin{equation}
\sin\theta=\frac{2}{\sqrt{3}}\sin\theta_{W}\label{Eq.10}\end{equation}

\paragraph{Boson mass spectrum}

Using Eq. (7) one can express the masses of the non-diagonal bosons.
They are (according to the parameter order in matrix $\eta^{2}$):

\begin{equation}
m_{W}^{2}=m^{2}a\label{Eq.11}\end{equation}

\begin{equation}
m_{X}^{2}=m^{2}\left[1-\frac{1}{2}(a+b)\right]\label{Eq.12}\end{equation}

\begin{equation}
m_{Y}^{2}=m^{2}\left[1-\frac{1}{2}(a-b)\right]\label{Eq.13}\end{equation}

Throughout this paper we consider $m^{2}=g^{2}\left\langle \phi\right\rangle ^{2}(1-\eta_{0}^{2})/4$. 

Evidently, $W$ is the ''old'' charged boson of the SM which links
positions 2 - 3 in the fermion triplet, namely the left-handed neutrino
to its charged lepton partner and, respectively, the ''up'' quarks
to ''down'' quarks. The neutral $Y$ boson couples the left-handed
neutrino to the right-handed one, and the ''classical'' up (down)
quarks to the ''exotic'' up (down) quarks, that is positions 1 -
2 in fermion triplet are involved. The remaining $X$ boson is responsable
for the charged current between positions 1 - 3 in each triplet. 

The ''pure'' neutral bosons (diagonal ones) aquire mass by diagonalizing
the resulting matrix:

\begin{equation}
M^{2}=m^{2}\left(\begin{array}{ccc}
1-\frac{1}{2}a+\frac{1}{2}b &  & -\frac{1}{\sqrt{3-4s^{2}}}\left(1-\frac{3}{2}a-\frac{1}{2}b\right)\\
\\-\frac{1}{\sqrt{3-4s^{2}}}\left(1-\frac{3}{2}a-\frac{1}{2}b\right) &  & \frac{1}{3-4s^{2}}\left(1+\frac{3}{2}a-\frac{3}{2}b\right)\end{array}\right)\label{Eq.14}\end{equation}
 after combining Eqs. (8), (9) and (6), where the notation $\sin\theta_{W}=s$
has been made for simplicity. 

One of the two diagonal bosons has to be identical to the neutral
boson $Z$ from SM. Therefore, the latter should be an eigenvector
of this mass matrix corresponding to the eigenvalue $m_{Z}^{2}=m_{W}^{2}/\cos^{2}\theta_{W}$
firmly established in the SM \cite{key-1}. 

\begin{equation}
M^{2}\mid Z>=\frac{m^{2}a}{1-s^{2}}\mid Z>\label{Eq.15}\end{equation}

That is, one computes $Det\left|M^{2}-m^{2}a/(1-s^{2})\right|=0$
which leads to the constraint upon the parameters $b=a\tan^{2}\theta_{W}$.
Consequently, the parameter matrix (6) becomes:

\begin{equation}
\eta^{2}=\left(1-\eta_{0}^{2}\right)diag\left[1-a,\frac{a}{2\cos^{2}\theta_{W}},\frac{a}{2}(1-\tan^{2}\theta_{W})\right]\label{Eq.16}\end{equation}

Under these circumstances, the boson mass spectrum yields:

\begin{equation}
m_{W}^{2}=m^{2}a\label{Eq.17}\end{equation}

\begin{equation}
m_{X}^{2}=m^{2}\left(1-\frac{a}{2\cos^{2}\theta_{W}}\right)\label{Eq.18}\end{equation}

\begin{equation}
m_{Y}^{2}=m^{2}\left[1-\frac{a}{2}(1-\tan^{2}\theta_{W})\right]\label{Eq.19}\end{equation}

\begin{equation}
m_{Z}^{2}=\frac{m^{2}a}{\cos^{2}\theta_{W}}\label{Eq.20}\end{equation}

\begin{equation}
m_{Z^{\prime}}^{2}=m^{2}\left[1+\frac{1}{3-4\sin^{2}\theta_{W}}-a\left(1+\frac{\tan^{2}\theta_{W}}{3-4\sin^{2}\theta_{W}}\right)\right]\label{Eq.21}\end{equation}
since $Tr(M^{2})=m_{Z}^{2}+m_{Z^{\prime}}^{2}$ holds. 

We obtained a mass spectrum depending on the free parameter $a$ (to
be tuned). One can observe that, although the fermion representations
and even the order in the parameter matrix $\eta^{2}$ are not the
same with those choosen in Ref. \cite{key-3}, the resulting mass
spectrum has the same structure. That means there are equivalent ways
to chose the parameters in the general method in order to reach the
same particle content of the model and the same physics.

\section{Critical Point}

When inspecting the boson mass spectrum - Eqs. (17) - (21) - one can
enforce certain conditions on the parameter $a$ as to obtain realistic
values, in accordance with the available experimental data. Furthermore,
the neutrino phenomenology was investigated \cite{key-4} and, because
a very high breaking scale $\left\langle \phi\right\rangle $ was
required, the method led to a natural see-saw mechanism \cite{key-5}.
It was thus implemented in order to keep consistency of the tiny masses
of neutrinos (and their specific mixing angles) with the reasonable
masses of the new bosons. Mass values in the region of TeVs \cite{key-1}
were allowed for the new bosons, since this see-saw mechanism was
exploited by inserting a supplemental small parameter in the $\eta^{2}$
matrix. 

However, a special and unexplored yet opportunity is offered by our
method. As long as the exact masses of the new bosons have not been
experimentaly determined to date, one is entitled to ask if there
is no screening between them. Namely, if the new neutral boson does
not ''cover'' the old one. More specifically, if their masses do
not coincide? What kind of consequnces has such a hypothesis?

From Eqs. (20) and (21) results that the free parameter has to be

\begin{equation}
a=\frac{2\cos^{2}\theta_{W}}{\mathbf{3-2\sin^{2}\theta_{W}}}\label{Eq.22}\end{equation}
in order to achieve $m_{Z}=m_{Z^{\prime}}$. That is $a\simeq0.6$
if we consider $\sin^{2}\theta_{W}\simeq0.223$. 

Furthermore, what are the values gained by the masses of the remaining
bosons? Embeding (22) in (17), (18) and (19) respectively, one obtains
the amazing results:

\begin{equation}
m_{W}^{2}=m_{X}^{2}=m^{2}\frac{2\cos^{2}\theta_{W}}{\mathbf{3-2\sin^{2}\theta_{W}}}\label{Eq.23}\end{equation}
and

\begin{equation}
m_{Z}^{2}=m_{Z^{\prime}}^{2}=m_{Y}^{2}=m^{2}\frac{2{}}{\mathbf{3-2\sin^{2}\theta_{W}}}\label{Eq.24}\end{equation}
.

These are the well-known values predicted by SM, namely $91.2$GeV
for the neutral bosons, and $80.4$ GeV for the charged ones. Assuming
that in the SM 

\begin{equation}
m_{W}=\frac{g}{2}\left\langle \phi\right\rangle _{SM}\label{Eq.25}\end{equation}
holds, one can estimate the required breaking scale $\left\langle \phi\right\rangle $
of the 3-3-1 model under consideration here, by comparing it to (17).
That is $\left\langle \phi\right\rangle \geq\left\langle \phi\right\rangle _{SM}/\sqrt{a}$.
This leads to $\left\langle \phi\right\rangle \geq320$GeV.

\section{Concluding remarks}

We have proven in this brief report that the exact algebraical approach
for solving gauge models with high symmetries offers - when it is
applied to a 3-3-1 model with right-handed neutrinos - a plausible
explanation for the undescovered yet new bosons predicted by such
a model. They could well be ''screened'' by the ''old'' bosons,
since a particular critical value is assigned to the free parameter
$a$ of the method. This can occur at a not very high breaking scale
$\left\langle \phi\right\rangle \geq320$GeV, while the whole SM content
is naturally recovered (as it was shown in Ref. \cite{key-7}).

Now, if one wants to deal with the correct neutrino phenomenology,
one has to adjust the above solution, since such a value for the free
parameter - $a\simeq0.6$ - leads to an unacceptable order of magnitude
for the individual masses in the neutrino sector where some tensor-like
products between Higgs triplets are employed in the generating mass
terms \cite{key-3,key-4}. There are several ways out of this unnatural
outcome in order to accomodate the small neutrino masses in this model:
(a) one can resort to the well-known see-saw mechanism (as in Ref.
\cite{key-5}) by adding a small new parameter in the matrix $\eta^{2}$
without spoiling these results, or (b) one can impose certain supplemental
global symmetries (like, for instance, $L_{e}-L_{\mu}-L_{\tau}$)
\cite{key-6} where also a $\mu-\tau$ interchange symmetry can be
invoked, or (c) one can even conceive a suitable radiative mechanism
like in Refs. \cite{key-13,key-25,key-33} that gives rise to the
neutrino masses only at one-loop or two-loop level. 

Therefore, we consider that the strange coincidence that simultanously
occurs - namely, $m_{W}^{2}=m_{X}^{2}$ and $m_{Z}^{2}=m_{Z^{\prime}}^{2}=m_{Y}^{2}$
- for a particular value of the free parameter seems more than a simple
''fit''. It seems to express a possible deeper identity between
the ''same charge'' bosons. This hypothesis can not be ruled out
\emph{a priori}, since more accurate results regarding the decays
of the ''new'' bosons and high-energy scatterings involving their
couplings to fermions - experimental details which can reveal some
restrictions on the parameter $a$ - have to be more exactly investigated
at LHC.

\end{document}